\documentclass{article}

\usepackage{microtype}
\usepackage{graphicx}
\usepackage{subfigure}
\usepackage{booktabs} % for professional tables
\usepackage{enumitem}

\usepackage{hyperref}

\usepackage[accepted]{icml2020}

\icmltitlerunning{Facial Recognition: A cross-national Survey on Public Acceptance, Privacy, and Discrimination}

\begin{document}

\twocolumn[
\icmltitle{Facial Recognition: A cross-national Survey on Public Acceptance, \\Privacy, and Discrimination}

%%%%%%%%%%%%%%%%%%%%%%%%%%%%%%%%%%%%%%%%%%%%%%%%%%%%%%%%%%%%%%%%
%
%  Authors


\begin{icmlauthorlist}
\icmlauthor{L{\'e}a Steinacker}{1}
\icmlauthor{Miriam Meckel}{1}
\icmlauthor{Genia Kostka}{2}
\icmlauthor{Damian Borth}{1}

\end{icmlauthorlist}

\icmlaffiliation{1}{University of St.Gallen, Switzerland}
\icmlaffiliation{2}{Freie Universit{\"a}t Berlin, Germany}

\icmlcorrespondingauthor{L{\'e}a Steinacker}{lea.steinacker@student.unisg.ch}

%   Important: these are used for review selection
%
\icmlkeywords{Face Recognition, Survey, Privacy, Trust, Surveillance, Legal, Policy, Discrimination}
\vskip 0.3in
]



\printAffiliationsAndNotice{}  % leave blank if no need to mention equal contribution

%%%%%%%%%%%%%%%%%%%%%%%%%%%%%%%%%%%%%%%%%%%%%%%%%%%%%%%%%%%%%%%%
%
%   Abstract!!!
%   Check check:
%   - face recognition vs. facial recognition
%   - using the post-fix "technology"
%   -> decide which one to use consistently in the paper
%
%%%%%%%%%%%%%%%%%%%%%%%%%%%%%%%%%%%%%%%%%%%%%%%%%%%%%%%%%%%%%%%%%

\begin{abstract}
With rapid advances in machine learning (ML), more of this technology is being deployed into the real world interacting with us and our environment. One of the most widely applied application of ML is facial recognition as it is running on millions of devices. While being useful for some people, others perceive it as a threat when used by public authorities. This discrepancy and the lack of policy increases the uncertainty in the ML community about the future direction of facial recognition research and development.
In this paper we present results from a cross-national survey about public acceptance, privacy, and discrimination of the use of facial recognition technology (FRT) in the public. This study provides insights about the opinion towards FRT from China, Germany, the United Kingdom (UK), and the United States (US), which can serve as input for policy makers and legal regulators.
\end{abstract}

%%%%%%%%%%%%%%%%%%%%%%%%%%%%%%%%%%%%%%%%%%%%%%%%%%%%%%%%%%%%%%%%
%
%   Introduction
%   - more and more face recognition systems are deployed
%   - people are mixed - with tendency towards negative sentiment
%   - conclusion we need a consensus about how people feel regarding FaceRec
%
%%%%%%%%%%%%%%%%%%%%%%%%%%%%%%%%%%%%%%%%%%%%%%%%%%%%%%%%%%%%%%%%%
\section{Introduction}
\label{introduction}
Deep learning has been able to demonstrate great potential in many domains such as computer vision, natural language processing or times series analysis with the consequence of deep neural networks becoming the core technology of many products and services. One of the most successful machine learning applications deployed into the real world is \textit{facial recognition} as it is currently running on millions of mobile phones for access control. As major international companies like Apple, Facebook, Google, Alibaba, and Baidu have integrated facial recognition into their platforms for recreational purposes, an ever-expanding trove of data is now available for processing to detect and identify billions of human faces. 

In addition to its myriad applications on personal devices, facial recognition technology (FRT) is used by law enforcement agencies around the world to monitor the public space via biometric data collection. In particular in this context, researchers have voiced concerns of FRT's discriminatory effects \cite{braca_investigation_2017, ngan_face_2015}, racial bias in accuracy \cite{buolamwini_gender_2018}, flawed data \cite{garvie_garbage_2019}, and privacy violations \cite{milligan1999facial, schwartz2012chicago}. This debate has spurred calls for greater accountability and oversight of FRT \cite{mann_automated_2017, naker_now_2017}.

While some local US governments have banned the use of FRT by city and state agencies, there is currently no federal legislative consensus despite extensive activism for regulation. Similarly, neither the European Commission nor any of its member states have explicitly ruled on FRT. However, the General Data Protection Regulation (GDPR) does protect and require consent for the collection of personal data, particularly sensitive information about an identifiable living individual, and some nations have specified it to include biometrics. It remains unclear though if facial images always fall under the GDPR's scope, depending, for example, on the legal justification for processing because substantial public interest such as national security or public safety may afford a path for circumventing consent \cite{buckley2011say}. In China, FR technology fused with other big data collection tools have become central to the government’s plan to be the world’s leader in artificial intelligence. 

Evidently, FRT and the law interact in two important ways: First, the technology directly serves purposes of law enforcement with consequences for citizens around the world. Second, as regulation often lags behind technological innovation, laws to systematically manage the application of FRT in the private and public realm are currently largely lacking.  

Amidst global protests against police brutality in June 2020, Amazon announced a year-long moratorium on police use of its widespread FRT software Rekognition, and Microsoft subsequently pledged not to sell their technology to US police departments until there is national regulation. IBM announced the discontinuation of their development and deployment of FRT altogether \cite{magid_ibm_nodate}. However, companies like Clearview AI continue to serve law enforcement around the world with its unprecedented facial recognition database of more than three billion images scraped from major online platforms \cite{hill2020secretive}. Such influential decisions of multinational technology companies do not only fill the regulatory vacuum but may also impact the future trajectory of FRT.

Given its growing global usage and despite scarce regulatory consensus so far, the question of \textit{what drives international public acceptance} of FRT is of timely importance and of value to inform the discussion of policy. In this paper, we focus on people's impressions of various policy-relevant dimensions and their effects on acceptance of FRT used in public across four countries: China, Germany, the United Kingdom (UK), and the United States (US). Our contribution is threefold:

\begin{enumerate}[topsep=0pt,itemsep=0.5ex,partopsep=1ex,parsep=1ex]

    \item We present survey results of a cross-national study on attitudes towards FRT used in public
    \item We demonstrate the impact of policy-relevant concerns such as privacy and discrimination on acceptance of FRT
    \item We derive conclusions that serve the ML and law communities
    
\end{enumerate}

The remainder of this paper is structured as follows. In the next section we present work related to facial recognition advances, its use for law enforcement purposes, and research on public approval. We then describe the methodology and partial results of our cross-national survey. Finally, we discuss our findings and synthesize conclusions.  

%%%%%%%%%%%%%%%%%%%%%%%%%%%%%%%%%%%%%%%%%%%%%%%%%%%%%%%%%%%%%%%%
%
%   State-of-the-art / Related Work
%   - Damian: face recognition from ML perspective, technical maturity
%   - Damian: discussion about this tech in the ML community
%   - overview of surveys in Face Recognition domain regarding policy
%   - question: How do policy dimensions impact public acceptance of facial recognition technology? 
%
%%%%%%%%%%%%%%%%%%%%%%%%%%%%%%%%%%%%%%%%%%%%%%%%%%%%%%%%%%%%%%%%%
\section{Related Work}
\label{sec:related_work}

%
% ml sota -> survey and overview 
% adversarial attack -> false positives - relevant
% deep fake - not that much
% face recognition critics: bias, fairness -> might not be relevant that much
%
\subsection{Robustness and Vulnerability of Facial Recognition}
In machine learning, the development of facial recognition technology has a rich history covering early days of face detection~\cite{viola2004robust} until the use of deep neural networks~\cite{parkhi2015a} for robust face recognition. An important component fueling the development of more robust systems was the availability of datasets and  defined performance evaluation metrics~\cite{phillips2005overview}. Initially, datasets were curated to represent different poses and illumination conditions~\cite{schroff2015facenet} to evaluate recognition robustness from a computer vision point of view. But research effort quickly moved from controlled conditions to the use of face recognition in the wild
~\cite{huang2008labeled, huang2014labeled}, where accuracies of more than $92\%$ could be achieved~\cite{zhu2015high}.

Almost in parallel research about the vulnerability of facial recognition was able to demonstrate significant challenges for the robustness of such systems~\cite{kose2013vulnerability,scherhag2019face}. The robustness of face recognition systems was even more challenges with the rise of adversarial attacks~\cite{sharif2016accessorize} and the lack of explainability or interpretability of deep neural networks~\cite{goswami2018unravelling}. In particular~\cite{sharif2016accessorize} was able to show that face recognition system can be explicitly confused by simple adversarial attacks to impersonate people cross-gender and cross-race i.e. confuse and impersonate a South-Asian female with a Middle-Eastern male.

\subsection{In the Face of Law Enforcement}
Like with other forms of surveillance technology, increasing applications of FRT by law enforcement have prompted debates about the balancing of privacy rights and security measures \cite{mccoy2002big}. Legal scholars who consider existing privacy law insufficiently suited for FRT that may scan publicly available images without requiring consent of individuals have argued for the need of a user opt-out regime \cite{mcclurg2007face}. In combination with live-tracking and body-worn cameras, some posit the use of FRT by law enforcement might also redefine public spaces by erasing anonymity therefore endangering free speech protections \cite{ringrose2019law}.

Concerns about the technology's variations in accuracy are particularly noteworthy when FRT is expected to produce reliable, admissible evidence to be used for law enforcement. Studies have demonstrated that bias in performance disproportionately impacts women and people of color \cite{klare2012face} as well as transgender and non-binary persons \cite{scheuerman2019computers}. First cases were already reported where a person was wrongfully accused by FRT\footnote{\url{https://www.nytimes.com/2020/06/24/technology/facial-recognition-arrest.html}}.

% https://www.nytimes.com/2020/06/24/technology/facial-recognition-arrest.html

Given the high stakes of consequences for already vulnerable populations, potential abuse of FRT is another concern. Controversial research efforts to demonstrate a correlation between facial features and criminality \cite{wu2016automated} or sexual orientation \cite{wang2018deep} signify the possibilities for misappropriating the technology, such as applying FRT for tracking ethnic minorities \cite{leibold2020surveillance}. 

\subsection{Public Recognition}

Previous studies have shown that people are most likely to accept technologies, including facial recognition, that they are most familiar with \cite{buckley2019language}. This suggests prevalence of FRT may affect familiarity and thus acceptance. International implementation of FRT by both governments and private companies vary widely. In China, the administration's agenda for AI expansion has pushed FRTs to the forefront of governmental surveillance and commercial smartphone applications are widespread. Similarly in the UK, a nation long equipped with extensive CCTV particularly in its capital, both law enforcement actors and private companies have employed FRT. In the US, public and commercial use of FRT has become increasingly widespread. While the 9/11 terror attacks prompted a surge in the implementation of technology-driven public safety measures including FRT, major US platform companies' integration of the technology has rapidly expanded facial databases. In Germany, by contrast, two historical precedents of oppressive surveillance states in the recent past form the backdrop to FRT adoption that has been lower in terms of speed and scope.

Research of FRT acceptance by the public has largely focused nationally. While no cross-cultural study has been done in the four countries investigated in our focus, approval rates of biometric surveillance technologies vary international comparison. 

In one study of Chinese Internet users, 80\% of participants either somewhat or strongly approved of the government's social credit surveillance toolbox, which includes FRT, with higher education predicting higher acceptance rates \cite{kostka2019china}. While studies have assessed public opinion of surveillance technologies by German citizens \cite{van2016surveillance}, none so far has concentrated on public opinion of FRT specifically. When a UK research institute surveyed 4,109 adults, 49\% support its use in policing practices given appropriate safeguards, but 67\% oppose it in schools, 61\% on public transport, and a majority of 55\% want restriction placed on its use by police \cite{ada_lovelace_institute_beyond_2019}. In the US, a Pew Research Center’s survey of 4,272 American adults showed that more than half trust law enforcement actors to employ FRT responsibly, but only 36\% think the same of technology companies, and 18\% of advertisers \cite{smith2019more}.

% Maybe still use: FRT challenges cultural notions of security versus civil liberties \cite{bowyer_face_2004, gray_urban_2003}. In the US alone, one in two American adults is in a database for FR by law enforcement \cite{garvie_perpetual_2016}.

%%%%%%%%%%%%%%%%%%%%%%%%%%%%%%%%%%%%%%%%%%%%%%%%%%%%%%%%%%%%%%%%
%
%   Methodology (Léa) 
%   - how was the study done
%   - what does the dataset look like
%
%%%%%%%%%%%%%%%%%%%%%%%%%%%%%%%%%%%%%%%%%%%%%%%%%%%%%%%%%%%%%%%%%
\section{Methodology}
\label{sec:methodology}
The results presented in this paper are based on an online survey we ran during August and September 2019 in China, Germany, the UK, and the US. A full version of the survey results can be found in~\cite{kostka2020eyes}.

\subsection{Study Design}
%During August and September 2019, we ran a cross-national online survey in China, Germany, the UK, and the US. 
Executed by a Berlin-based firm who cooperates with providers in each of the four countries, the survey employed a river sampling method, drawing both first-time and regular survey participants from a base of 1-3 million unique online users through mobile applications. 
Our survey reached respondents through more than 100 apps out of a network of over 40,000 participating partners. 
Formats and genres of the applications vary, including e-commerce, photo-sharing, and messaging. 
Within each application, offer walls provided participants options to receive small financial and non-monetary rewards, such as premium content, extra features, vouchers, and PayPal cash in exchange for taking part in the survey without knowing the topic before opting in. After an initial pre-screening that matched participants with a survey, the conversion rate of users who fully finished ours was $70\%$ (China), $73\%$ (Germany), $69\% (UK)$, and $67\% (US)$ respectively. Consecutive identical answer choices or disproportionately short periods of time for completion of a questionnaire were reasons for invalidation. Our sample comprised $6,633$ respondents from all four countries in total, sampled based on age (18-65), gender and region. Given the nature of conducting the survey online, this sample resembles a nationally representative group of the Internet-connected population: likely slightly younger and more technology-savvy than the overall population. Collected data was weighted by each aforementioned variable with a maximum weight of $1.8.$ \footnote{Given an estimated design effect of $1.03$ for China, $1$ for Germany, $1.06$ for the UK, and $1.04$ for the US, the overall margin of error for estimates is $2.4\%$ for China, $2.4\%$ for Germany, $2.5\%$ for the UK, and $2.5\%$ for the US.}

\subsection{Data Analysis}
Of the $6,633$ respondents in our sample, $8.1\% (N=534)$ had \emph{“never heard about FRT”} prior to taking the survey. As their attitudes are nonetheless relevant for understanding overall public opinion on the matter, we did not exclude their further responses. After the initial gauge of awareness, a short prompt summarized for participants: \emph{"Facial recognition technology is used to automatically identify people by scanning their face from an image or video."} For the purposes of this paper, we sought to understand, first, participants' stance on FR deployment in the public sphere (\emph{"Do you accept or oppose the use of facial recognition technology in public?”}) and, second, how they interpreted related issues. We ran ordered logit regressions with the dependent variable of interest being ``social acceptance'' to investigate the effects of privacy threat impressions, trust in government, and perceptions of surveillance on participants’ acceptance levels controlling for age, gender, and education.  

%%%%%%%%%%%%%%%%%%%%%%%%%%%%%%%%%%%%%%%%%%%%%%%%%%%%%%%%%%%%%%%%
%
%   Results
%   - how was the study done
%   - how is the dataset looking like
%
%%%%%%%%%%%%%%%%%%%%%%%%%%%%%%%%%%%%%%%%%%%%%%%%%%%%%%%%%%%%%%%%%

\section{Results}
\label{sec:results}

%The results of our survey are presented according to ...

\subsection{Social Acceptance}
Overall, in the Chinese sample, considerably more participants accept than oppose FRT use in public, while in the UK, only slightly more do so. On the contrary, among the German and  U.S. respondents, slightly more oppose than accept it.  

%   At Damian: Die folgende Tabelle wäre als horizontaler Bar Graph gut.
 
 \begin{table}[h]
 \caption{\textit{Results of survey question: ``Do you accept or oppose the use of FRT in public?''}}
 \label{tab:acceptance}
 \vskip 0.05in
 \begin{center}
 \begin{scriptsize}
 \begin{sc}
 \begin{tabular}{lcccr}
 \toprule
 Response &  China & Germany & UK & US \\
 \midrule
  Strongly oppose    & 4\% & 18\% & 14\% & 16\% \\
 Somewhat oppose & 18\% & 21\% & 19\% & 22\% \\
 Neither oppose nor accept   & 28\% & 24\% & 27\% & 27\% \\
 Somewhat accept    & 39\% & 29\% & 28\% & 24\% \\
 Strongly accept     & 11\% & 8\% & 11\% & 11\% \\
 \bottomrule
 \end{tabular}
 \end{sc}
 \end{scriptsize}
 \end{center}
 \vskip -0.05in
 \end{table}

Specifically, as Table \ref{tab:acceptance} shows, 50\% of the respondents from China, 37\% from Germany, 39\% from the UK, and 35\% from the US \textit{somewhat} or \textit{strongly accept} the use of FRT in public. Meanwhile, 22\% from China, 39\% from Germany, 33\% from the UK, and 38\% from the US respondents \textit{somewhat} or \textit{strongly oppose} it.

\subsection{Impressions of FRT}
To understand contributing factors of participants' acceptance levels, we examined their impressions of relevant dimensions related to FRT. First, we gauged perception of FRT's consequences, such as \textit{privacy violations}, \textit{discrimination}, and \textit{surveillance}, as well as \textit{convenience}, \textit{efficiency}, and \textit{security}.

 \begin{table}[h]
 \caption{Results of survey question: \textit{``Do you think FRT increases any of the following?''}}
 \label{tab:increase}
 \vskip 0.05in
 \begin{center}
 \begin{scriptsize}
 \begin{sc}
 \begin{tabular}{lcccr}
 \toprule
 Response &  China & Germany & UK & US \\
 \midrule
  Privacy violations & 31\% & 48\% & 39\% & 44\% \\
  Surveillance    & 27\% & 62\% &53\% & 52\% \\
  Discrimination    & 3\% & 15\% & 16\% & 16\% \\
  Convenience   & 65\% & 22\% & 20\% & 30\% \\
  Efficiency  & 56\% & 20\% & 25\% & 31\% \\
  Security     & 62\% & 54\% &62\% & 64\% \\
  None of the above     & 3\% & 5\% & 8\% & 6\% \\
 \bottomrule
 \end{tabular}
 \end{sc}
 \end{scriptsize}
 \end{center}
 \vskip -0.05in
 \end{table}

As Table \ref{tab:increase} shows, almost half of all German and US respondents (48\% and 44\% respectively) expect FRT to increase privacy violations, while only 31\% of Chinese and 39\% of UK participants do. A clear majority of Germans (62\%) and roughly half of UK and US respondents believe that FRT increases surveillance, yet only 27\% of Chinese agree. Given the extensive research demonstrating various discriminatory effects of FRT (ref), it is notable that, in all four countries, very few participants, less than 1 in 5, consider FRT to be an exacerbating factor in discrimination. A majority of the participants from China (65\%) yet only 22\% in Germany, 20\% in the UK and 30\% in the US think that FRT increases convenience. More than half of Chinese respondents expect FRT to advance efficiency (56\%), while 1 in 5 of German, 1 in 4 of UK, and 1 in 3 of American respondents do. Strikingly, a majority in all four countries (62\%, 54\%, 62\%, and 64\% respectively) expect FRT to increase security. 

Second, given the context of FRT employed by law enforcement for official identification, we examined participants' impressions of the technology's \textit{reliability} in comparison to previous methods.  

 \begin{table}[h]
 \caption{Results of survey question: \textit{``Do you think that facial recognition technology is more reliable or less reliable than other identification methods (e.g.: fingerprints, identity cards)?''}}
 \label{tab:reliability}
 \vskip 0.05in
 \begin{center}
 \begin{scriptsize}
 \begin{sc}
 \begin{tabular}{lcccr}
 \toprule
 Response &  China & Germany & UK & US \\
 \midrule
  More reliable   & 43\% & 31\% & 32\% & 34\% \\
  Neither more nor less & 40\% & 38\% & 34\% & 37\% \\
  Less reliable  & 7\% & 19\% & 17\% & 15\% \\
  Don't know    & 10\% & 13\% & 12\% & 13\% \\
 \bottomrule
 \end{tabular}
 \end{sc}
 \end{scriptsize}
 \end{center}
 \vskip -0.05in
 \end{table}

As Table \ref{tab:reliability} illustrates, in all countries, fewer than 1 in 5 respondents rate FRT's reliability lower than other forms of identification, though China is the only country where a majority of participants perceive it as actually more reliable.  
  
\subsection{Concerns, Trust, Surveillance}
For further examination, we analyzed various dimensions relevant to policy-making for FRT: \textit{Issue concerns} involving law enforcement, levels of governmental \textit{trust}, and perceptions and support of \textit{surveillance} more generally. Across all four countries, a majority of participants is worried about crime (59\%, 67\%, 74\%, 74\%) and, except for China, about terrorist threats (47\%, 56\%, 66\%, 67\%). As Table \ref{tab:concerns} shows, the UK and the US stand out with over 40\% of their participants respectively indicating that they are concerned about the control of their nations' borders, while less than 30\% in both China and Germany do. Around half of Chinese respondents are concerned about violations of rules \& regulations, while around 38\% of Germans, 36\% of UK and 41\% of US participants are. Finally, a majority in both China (51\%) and the UK (54\%) find socially unacceptable behavior concerning. 

 \begin{table}[h]
 \caption{Results of survey question: \textit{``Are you concerned about any of the following issues in your country?''}}
 \label{tab:concerns}
 \vskip 0.05in
 \begin{center}
 \begin{tiny}
 \begin{sc}
 \begin{tabular}{lcccr}
 \toprule
 Response &  China & Germany & UK & US \\
 \midrule
  Crime & 59\% & 67\% & 74\% & 74\% \\ 
  Terrorist threats  & 47\% & 56\% & 66\% & 67\% \\
  Border control   & 29\% & 28\% & 42\% & 43\% \\
  Violations of rules \& regulations  & 49\% & 38\% & 36\% & 41\% \\
  Socially unacceptable behavior   & 51\% & 42\% & 54\% & 36\% \\
  None of the above   & 14\% & 14\% & 9\% & 10\% \\
 \bottomrule
 \end{tabular}
 \end{sc}
 \end{tiny}
 \end{center}
 \vskip -0.05in
 \end{table}

Linking the previous question into the context of trusting the local government (Table \ref{tab:trust}), we observe that with the exception of China ($7\%$), that people in Germany ($37\%$), the UK ($48\%$), and the US ($40\%$) only trust their government \textit{very little} or \textit{not at all}. On the contrary, while a majority of $61\%$ in China trust their administration \textit{a lot}, only $23\%$ in Germany, and merely $10\%$ in both the UK and the US do. 

 \begin{table}[h]
 \caption{Results of survey question: \textit{``How much do you trust governmental institutions in your country?''}}
 \label{tab:trust}
 \vskip 0.05in
 \begin{center}
 \begin{scriptsize}
 \begin{sc}
 \begin{tabular}{lcccr}
 \toprule
 Response &  China & Germany & UK & US \\
 \midrule
  Not at all & 1\% & 14\% & 17\% & 12\% \\
  Very little & 6\% & 23\% & 31\% & 28\% \\
  Somewhat  & 24\% & 32\% & 37\% & 42\% \\
  A lot   & 61\% & 23\% & 10\% & 10\% \\
  Prefer not to answer   & 7\% & 7\% & 5\% & 8\% \\
 \bottomrule
 \end{tabular}
 \end{sc}
 \end{scriptsize}
 \end{center}
 \vskip -0.05in
 \end{table}

Finally, we investigated participants' judgment of their government's domestic surveillance history and their current support of governmental surveillance. While $46\%$ of respondents in Germany, $41\%$ in the UK, and $54\%$ in the US believe their country's government has employed domestic surveillance negatively in the past, only $13\%$ of Chinese do (Table \ref{tab:surveillancehistory}). While in China and the UK, considerably more people somewhat or strongly support ($52\%$ and $44\%$) surveillance than somewhat or strongly oppose it ($16\%$ and $24\%$), in Germany ($34\%$ vs $40\%$) and the US ($31\%$ vs $37\%$) the two sides appear roughly equal. About a third of all respondents in each country neither oppose nor support surveillance in their country (Table \ref{tab:surveillancesupport}). 

 \begin{table}[t]
 \caption{Results of survey question: \textit{``Do you think that the government in your country has used surveillance against its own citizens in a negative way in the past?''}}
 \label{tab:surveillancehistory}
 \vskip 0.05in
 \begin{center}
 \begin{scriptsize}
 \begin{sc}
 \begin{tabular}{lcccr}
 \toprule
 Response &  China & Germany & UK & US \\
 \midrule
  Yes & 13\% & 46\% & 41\% & 54\% \\
  No & 37\% & 23\% & 19\% & 15\% \\
  Don't know  & 50\% & 30\% & 41\% & 32\% \\
 \bottomrule
 \end{tabular}
 \end{sc}
 \end{scriptsize}
 \end{center}
 \vskip -0.05in
 \end{table}

\subsection{Effects on Social Acceptance}
To gauge the effects of a number of these factors on public approval we ran an ordered logit regression with social acceptance of FRT use in public as our dependent variable and privacy threat perception, consequences of FRT, and national issue concerns as independent variables.  

Our analysis shows that the interpretation of privacy threat is a strong and significant negative predictor of acceptance (see Table~\ref{tab:ordered_logit_regression}). In other words, the more a participant perceives the technology as a risk to their privacy, the less likely they are to accept FRT use in public. This finding is statistically significant across all four countries and in each setting individually. 

Convenience has a significantly positive effect on acceptance overall, and in each country individually, with the exception of Germany, where the sign turns. That is to say, the more likely German participants perceive FRT to increase convenience, the less likely they are to accept it, pointing perhaps to cultural skepticism towards a hyped comfort of technology. 

Across all countries, overall and individually, impressions of increased efficiency as well as security raise the likelihood of acceptance as the corresponding coefficients are significant and positive. Discrimination is a significantly negative predictor overall as well as in Germany and the UK specifically while in China and the US, the coefficients are not statistically significant.  

In international comparison, the more a participant perceives FRT to increase surveillance the more likely they are to accept it, with the exception of China, where the effect is negative, and the US, where the result is not significant. 

Our analysis shows that the perception of terrorist threats is a significant positive predictor across all four countries. When perceived as a national concern, socially unacceptable behavior has a significant, positive relationship to approval of FRT in public overall and in Germany. On the other hand, neither concerns about violations of rules and regulations, nor about crime or border control are significantly linked with higher rates of acceptance.  

\begin{table}[t]
 \caption{Results of survey question: \textit{``Do you generally support or oppose the use of surveillance by your government in your country?''}}
 \label{tab:surveillancesupport}
 \vskip 0.05in
 \begin{center}
 \begin{scriptsize}
 \begin{sc}
 \begin{tabular}{lcccr}
 \toprule
 Response &  China & Germany & UK & US \\
 \midrule
  Strongly oppose & 4\% & 15\% & 9\% & 12\% \\
  Somewhat oppose & 12\% & 19\% & 15\% & 19\% \\
  Neither oppose nor support  & 32\% & 27\% & 31\% & 32\% \\
  Somewhat support   & 34\% & 32\% & 31\% & 26\% \\
  Strongly support   & 18\% & 8\% & 13\% & 11\% \\
 \bottomrule
 \end{tabular}
 \end{sc}
 \end{scriptsize}
 \end{center}
 \vskip -0.05in
 \end{table}

%%%%%%%%%%%%%%%%%%%%%%%%%%%%%%%%%%%%%%%%%%%%%%%%%%%%%%%%%%%%%%%%
%
%   Conclusion
%   - Léa: overall conclusions drawn from the results 
%   - Damian: relevance for the ML community 
%
%%%%%%%%%%%%%%%%%%%%%%%%%%%%%%%%%%%%%%%%%%%%%%%%%%%%%%%%%%%%%%%%%
\section{Conclusion}
\label{sec:conclusion}
How the scientific community continues to improve FRT and the law oversees it bears far-reaching implications for people worldwide. 
Public funding, young researchers in this area, and investments by tech companies are currently under pressure because of the uncertainty of what defines an \textit{acceptable} use of FRT. Such uncertainty can be resolved by guidelines from regulatory or legal entities. Such guidelines however work best when they reflect a consensus of the people's opinion about what an acceptable and valid use of FRT is. The machine learning community needs such guiding policies to reduce the current uncertainty towards future research. This is in particular important because of the challenges FRT is currently facing with regard to its vulnerability, adversarial attacks, or lack of interpretability or explainability.

The insights presented in this paper into some of the most salient factors influencing acceptance can inform the further development of applications and the formulation of policy. 
First, there appears to be no consensus on the public use of FRT across the four countries studied. Varying international acceptance levels signal that there might not be a feasible unified approach to governing FRT. Respondents from China and the UK express more acceptance than disavowal, compared to the opposite in Germany and the US. Similarly distributed are the frequencies of who perceives FR to increase privacy violations. When they do perceive the technology as a privacy threat, though, people across all countries are more likely to oppose its use in public. This finding suggests that a rise in reports and investigations of privacy violations related to FR might contribute to a decline in global acceptance in the future. 

Second, a majority of respondents in all four countries expect FR to increase security. This emphasizes that in the nations studied, the perceived main trade-off of FR appears to be between security and privacy.

Third, awareness of the discriminatory effects of applied FRT are very low across all four countries and our findings imply greater awareness would lead to lower acceptance. Given wide-ranging research showing their disproportionate impacts on vulnerable and minority populations, including women, people of color, and members of the LGBTQ community, this gap underlines the responsibility of developers to improve accuracy and biased performance and of law enforcement to ensure protection against discrimination, particularly through the use of FR in public. 

Fourth, across all countries studied a concern about terrorist threats affected acceptance rates more significantly than perceptions of crime or border control. This suggests that respondents have distinct thresholds of justification for FRT used by law enforcement. 

When applied with such consequential social ramifications, mitigating FRT's current discriminatory effects must be a focus of both the scientific and legislative communities. Furthermore, the scope of privacy protections must match technological advances and implementation purposes.

%
%   Large table - one page
%
%
\begin{table*}[t]
    \caption{\textit{Ordered Logit Regression, weighted, dependent variable: social acceptance of FRT in public}}
     \label{tab:ordered_logit_regression}
    \vskip 0.05in
    \begin{center}
    \begin{footnotesize}
    \begin{sc}
    %
    %   tab data
    %
    \begin{tabular}{lccccc}
    \toprule
    \multicolumn{6}{c}{\textbf{Ordered logit regression, weighted, dv: social acceptance of FRT (pubic)}} \\
    \midrule
    \textbf{}                  & \textbf{Total}          & \textbf{China}          & \textbf{Germany}        & \textbf{UK}            & \textbf{US}            \\
    \midrule
\textbf{Age}          & 0.00400**   & 0.00695     & 0.00042     & 0.00247     & 0.01023***  \\
\textbf{}             & (0.00180)   & (0.00484)   & (0.00358)   & (0.00339)   & (0.00366)   \\
\textbf{}                  & \multicolumn{1}{l}{}    & \multicolumn{1}{l}{}    & \multicolumn{1}{l}{}    & \multicolumn{1}{l}{}   & \multicolumn{1}{l}{}   \\
\textbf{Gender}       & -0.08896*   & 0.00598     & -0.20971**  & -0.10834    & 0.00739     \\
\textbf{}             & (0.04606)   & (0.09362)   & (0.09382)   & (0.09307)   & (0.09572)   \\
\textbf{Education}         & \multicolumn{1}{l}{}    & \multicolumn{1}{l}{}    & \multicolumn{1}{l}{}    & \multicolumn{1}{l}{}   & \multicolumn{1}{l}{}   \\
Medium                & 0.25815**   & 0.30059     & 0.51652***  & 0.00630     & -0.34853    \\
                      & (0.10351)   & (0.24525)   & (0.19545)   & (0.17221)   & (0.23558)   \\
High                  & 0.49870***  & 0.60922**   & 0.59164***  & 0.31901*    & -0.03548    \\
                      & (0.11129)   & (0.25556)   & (0.22217)   & (0.18750)   & (0.25031)   \\
    \midrule
\textbf{Privacy threat}    & \multicolumn{1}{l}{}    & \multicolumn{1}{l}{}    & \multicolumn{1}{l}{}    & \multicolumn{1}{l}{}   & \multicolumn{1}{l}{}   \\
    \midrule
Maybe                 & -1.19227*** & -0.88714*** & -1.47814*** & -1.25455*** & -1.13848*** \\
                      & (0.06499)   & (0.12911)   & (0.13558)   & (0.12816)   & (0.13783)   \\
Yes                   & -2.56136*** & -2.05615*** & -2.83184*** & -2.73087*** & -2.24377*** \\
                      & (0.09095)   & (0.19423)   & (0.17812)   & (0.19148)   & (0.17547)   \\
Don't know            & -1.10769*** & -0.61900*** & -1.65456*** & -1.14611*** & -1.14786*** \\
                      & (0.08928)   & (0.18081)   & (0.19379)   & (0.18921)   & (0.17015)   \\
    \midrule
\textbf{Consequences}      & \multicolumn{1}{l}{}    & \multicolumn{1}{l}{}    & \multicolumn{1}{l}{}    & \multicolumn{1}{l}{}   & \multicolumn{1}{l}{}   \\
    \midrule
Convenience           & 0.39866***  & 0.65575***  & -0.21672*   & 0.61789***  & 0.48929***  \\
                      & (0.05601)   & (0.10862)   & (0.12242)   & (0.13169)   & (0.11290)   \\
Privacy violations         & -0.47910***             & -0.30679***             & -0.51169***             & -0.53786***            & -0.50424***            \\
                      & (0.05638)   & (0.11384)   & (0.11292)   & (0.11722)   & (0.11150)   \\
Efficiency            & 0.38049***  & 0.27139***  & 0.26869**   & 0.48286***  & 0.48989***  \\
                      & (0.05647)   & (0.10358)   & (0.12690)   & (0.11988)   & (0.12034)   \\
Discrimination        & -0.39427*** & -0.15615    & -0.57354*** & -0.29403**  & -0.20140    \\
                      & (0.07769)   & (0.30014)   & (0.14208)   & (0.13921)   & (0.14378)   \\
Security              & 0.71487***  & 0.77954***  & 0.90220***  & 0.61509***  & 0.62676***  \\
                      & (0.05372)   & (0.10211)   & (0.10722)   & (0.12419)   & (0.11347)   \\
Surveillance          & 0.15870***  & -0.26568**  & 0.39048***  & 0.22459**   & 0.13438     \\
                      & (0.04934)   & (0.10949)   & (0.10508)   & (0.10463)   & (0.09962)   \\
None of the above     & 0.21330*    & 0.91277***  & 0.06282     & 0.22159     & 0.20214     \\
                      & (0.11358)   & (0.34577)   & (0.25441)   & (0.20607)   & (0.20929)   \\
    \midrule
\textbf{Issue concern}     & \multicolumn{1}{l}{}    & \multicolumn{1}{l}{}    & \multicolumn{1}{l}{}    & \multicolumn{1}{l}{}   & \multicolumn{1}{l}{}   \\
    \midrule
Violation of rules    & 0.06854     & 0.01662     & 0.15430     & -0.06003    & 0.10645     \\
and regulations       & (0.05191)   & (0.10302)   & (0.10442)   & (0.11323)   & (0.10658)   \\
Crime                 & -0.00350    & 0.01879     & 0.18941     & -0.15639    & 0.00426     \\
                      & (0.06169)   & (0.11331)   & (0.12497)   & (0.13409)   & (0.13524)   \\
Terrorist threats     & 0.27710***  & 0.05836     & 0.53743***  & 0.36463***  & 0.29392**   \\
                      & (0.05478)   & (0.11117)   & (0.10782)   & (0.11341)   & (0.12003)   \\
Border control        & 0.02710     & 0.05025     & 0.03798     & 0.09390     & 0.07452     \\
                      & (0.05257)   & (0.11220)   & (0.11135)   & (0.10449)   & (0.10537)   \\
Socially              & 0.17819***  & -0.08294    & 0.36925***  & 0.14626     & 0.07795     \\
unacceptable behavior & (0.05030)   & (0.10296)   & (0.10088)   & (0.11036)   & (0.11076)   \\
None of the above     & 0.16217*    & 0.00412     & 0.32852*    & -0.07788    & 0.35713*    \\
                      & (0.09183)   & (0.18155)   & (0.17772)   & (0.20161)   & (0.19948)   \\
    \midrule
    cut1                  & -2.45276*** & -2.84157*** & -2.28142*** & -2.83727*** & -2.32633*** \\
    Constant              & (0.15142)   & (0.35387)   & (0.29993)   & (0.29328)   & (0.32109)   \\
    cut2                  & -0.90108*** & -0.71077**  & -0.74262**  & -1.36520*** & -0.83437*** \\
    Constant              & (0.14857)   & (0.32728)   & (0.29838)   & (0.28760)   & (0.31508)   \\
    cut3                  & 0.54876***  & 0.76550**   & 0.70624**   & 0.15652     & 0.62404**   \\
    Constant              & (0.14808)   & (0.32551)   & (0.29969)   & (0.28672)   & (0.31392)   \\
    cut4                  & 2.70489***  & 3.11178***  & 3.17540***  & 2.27897***  & 2.40178***  \\
    Constant              & (0.15249)   & (0.33614)   & (0.31234)   & (0.29421)   & (0.32154)   \\
    \midrule
    Observations          & 6633        & 1651        & 1677        & 1685        & 1620        \\
    \midrule
    \multicolumn{6}{c}{\begin{tabular}[c]{@{}c@{}}Standard errors in parentheses\\ * p \textless 0.10, ** p \textless 0.05, *** p \textless 0.01\end{tabular}} \\
    \bottomrule
    \end{tabular}
    %
    %   tab data end
    %
    \end{sc}
    \end{footnotesize}
    \end{center}
    \vskip -0.05in
\end{table*}

%%%%%%%%%%%%%%%%%%%%%%%%%%%%%%%%%%%%%%%%%%%%%%%%%%%%%%%%%%%%%%%%
%
%   Acknowledgements - if we like
%   -> should only appear in the accepted version.
%
%%%%%%%%%%%%%%%%%%%%%%%%%%%%%%%%%%%%%%%%%%%%%%%%%%%%%%%%%%%%%%%%%
%\section*{Acknowledgements}

%\textbf{Do not} include acknowledgements in the initial version of the paper submitted for blind review.

%If a paper is accepted, the final camera-ready version can (and
%probably should) include acknowledgements. In this case, please
%place such acknowledgements in an unnumbered section at the
%end of the paper. Typically, this will include thanks to reviewers
%who gave useful comments, to colleagues who contributed to the ideas,
%and to funding agencies and corporate sponsors that provided financial
%support.

%%%%%%%%%%%%%%%%%%%%%%%%%%%%%%%%%%%%%%%%%%%%%%%%%%%%%%%%%%%%%%%%
%
%   Bib Tex File
%
%%%%%%%%%%%%%%%%%%%%%%%%%%%%%%%%%%%%%%%%%%%%%%%%%%%%%%%%%%%%%%%%%

% if you would like to do not cite
\nocite{langley00}
% standard setup to compile bib file
\bibliography{bibliography}

\begin{thebibliography}{36}
\providecommand{\natexlab}[1]{#1}
\providecommand{\url}[1]{\texttt{#1}}
\expandafter\ifx\csname urlstyle\endcsname\relax
  \providecommand{\doi}[1]{doi: #1}\else
  \providecommand{\doi}{doi: \begingroup \urlstyle{rm}\Url}\fi

\bibitem[Braca(2017)]{braca_investigation_2017}
Braca, A.
\newblock An investigation into {Bias} in {Facial} {Recognition} using
  {Learning} {Algorithms}.
\newblock 2017.

\bibitem[Buckley \& Hunter(2011)Buckley and Hunter]{buckley2011say}
Buckley, B. and Hunter, M.
\newblock Say cheese! privacy and facial recognition.
\newblock \emph{Computer Law \& Security Review}, 27\penalty0 (6):\penalty0
  637--640, 2011.

\bibitem[Buckley \& Nurse(2019)Buckley and Nurse]{buckley2019language}
Buckley, O. and Nurse, J.~R.
\newblock The language of biometrics: Analysing public perceptions.
\newblock \emph{Journal of Information Security and Applications}, 47:\penalty0
  112--119, 2019.

\bibitem[Buolamwini \& Gebru(2018)Buolamwini and Gebru]{buolamwini_gender_2018}
Buolamwini, J. and Gebru, T.
\newblock Gender shades: {Intersectional} accuracy disparities in commercial
  gender classification.
\newblock In \emph{Conference on fairness, accountability and transparency},
  pp.\  77--91, 2018.

\bibitem[Garvie(2019)]{garvie_garbage_2019}
Garvie, C.
\newblock Garbage {In}. {Garbage} {Out}. {Face} {Recognition} on {Flawed}
  {Data}, May 2019.
\newblock URL \url{https://www.flawedfacedata.com}.
\newblock Library Catalog: www.flawedfacedata.com.

\bibitem[Goswami et~al.(2018)Goswami, Ratha, Agarwal, Singh, and
  Vatsa]{goswami2018unravelling}
Goswami, G., Ratha, N., Agarwal, A., Singh, R., and Vatsa, M.
\newblock Unravelling robustness of deep learning based face recognition
  against adversarial attacks.
\newblock In \emph{Thirty-Second AAAI Conference on Artificial Intelligence},
  2018.

\bibitem[Hill(2020)]{hill2020secretive}
Hill, K.
\newblock The secretive company that might end privacy as we know it.
\newblock \emph{New York Times}, 18, 2020.

\bibitem[Huang \& Learned-Miller(2014)Huang and
  Learned-Miller]{huang2014labeled}
Huang, G.~B. and Learned-Miller, E.
\newblock Labeled faces in the wild: Updates and new reporting procedures.
\newblock \emph{Dept. Comput. Sci., Univ. Massachusetts Amherst, Amherst, MA,
  USA, Tech. Rep}, pp.\  14--003, 2014.

\bibitem[Huang et~al.(2008)Huang, Mattar, Berg, and
  Learned-Miller]{huang2008labeled}
Huang, G.~B., Mattar, M., Berg, T., and Learned-Miller, E.
\newblock Labeled faces in the wild: A database forstudying face recognition in
  unconstrained environments.
\newblock 2008.

\bibitem[Institute(2019)]{ada_lovelace_institute_beyond_2019}
Institute, A.~L.
\newblock Beyond face value: public attitudes to facial recognition technology,
  2019.
\newblock URL
  \url{https://www.adalovelaceinstitute.org/beyond-face-value-public-attitudes-to-facial-recognition-technology/}.
\newblock Library Catalog: www.adalovelaceinstitute.org.

\bibitem[Klare et~al.(2012)Klare, Burge, Klontz, Bruegge, and
  Jain]{klare2012face}
Klare, B.~F., Burge, M.~J., Klontz, J.~C., Bruegge, R. W.~V., and Jain, A.~K.
\newblock Face recognition performance: Role of demographic information.
\newblock \emph{IEEE Transactions on Information Forensics and Security},
  7\penalty0 (6):\penalty0 1789--1801, 2012.

\bibitem[Kose \& Dugelay(2013)Kose and Dugelay]{kose2013vulnerability}
Kose, N. and Dugelay, J.-L.
\newblock On the vulnerability of face recognition systems to spoofing mask
  attacks.
\newblock In \emph{2013 IEEE International Conference on Acoustics, Speech and
  Signal Processing}, pp.\  2357--2361. IEEE, 2013.

\bibitem[Kostka(2019)]{kostka2019china}
Kostka, G.
\newblock China’s social credit systems and public opinion: Explaining high
  levels of approval.
\newblock \emph{New media \& society}, 21\penalty0 (7):\penalty0 1565--1593,
  2019.

\bibitem[Kostka et~al.(2020)Kostka, Steinacker, and Meckel]{kostka2020eyes}
Kostka, G., Steinacker, L., and Meckel, M.
\newblock {In the Eyes of Citizens: Attitudes Towards Facial Recognition
  Technology in China, Germany, the UK and the US}.
\newblock \emph{Available at SSRN: https://ssrn.com/abstract=3518857}, 2020.

\bibitem[Leibold(2020)]{leibold2020surveillance}
Leibold, J.
\newblock Surveillance in china’s xinjiang region: Ethnic sorting, coercion,
  and inducement.
\newblock \emph{Journal of Contemporary China}, 29\penalty0 (121):\penalty0
  46--60, 2020.

\bibitem[Magid(2020)]{magid_ibm_nodate}
Magid, L.
\newblock {IBM}, {Microsoft} {And} {Amazon} {Not} {Letting} {Police} {Use}
  {Their} {Facial} {Recognition} {Technology}, 2020.
\newblock URL
  \url{https://www.forbes.com/sites/larrymagid/2020/06/12/ibm-microsoft-and-amazon-not-letting-police-use-their-facial-recognition-technology/}.
\newblock Library Catalog: www.forbes.com Section: Innovation.

\bibitem[Mann \& Smith(2017)Mann and Smith]{mann_automated_2017}
Mann, M. and Smith, M.
\newblock Automated facial recognition technology: {Recent} developments and
  approaches to oversight.
\newblock \emph{UNSWLJ}, 40:\penalty0 121, 2017.

\bibitem[McClurg(2007)]{mcclurg2007face}
McClurg, A.~J.
\newblock In the face of danger: Facial recognition and the limits of privacy
  law.
\newblock \emph{Harvard Law Review}, 120\penalty0 (7):\penalty0 1870--1891,
  2007.

\bibitem[McCoy(2002)]{mccoy2002big}
McCoy, S.
\newblock O'big brother where art thou?: The constitutional use of
  facial-recognition technology.
\newblock \emph{Order}, 20, 2002.

\bibitem[Milligan(1999)]{milligan1999facial}
Milligan, C.~S.
\newblock Facial recognition technology, video surveillance, and privacy.
\newblock \emph{S. Cal. Interdisc. LJ}, 9:\penalty0 295, 1999.

\bibitem[Naker \& Greenbaum(2017)Naker and Greenbaum]{naker_now_2017}
Naker, S. and Greenbaum, D.
\newblock Now you see me: {Now} you still do: {Facial} recognition technology
  and the growing lack of privacy.
\newblock \emph{BUJ Sci. \& Tech. L.}, 23:\penalty0 88, 2017.

\bibitem[Ngan et~al.(2015)Ngan, Grother, and Ngan]{ngan_face_2015}
Ngan, M., Grother, P.~J., and Ngan, M.
\newblock \emph{Face recognition vendor test ({FRVT}) performance of automated
  gender classification algorithms}.
\newblock US Department of Commerce, National Institute of Standards and
  Technology, 2015.

\bibitem[Parkhi et~al.(2015)Parkhi, Vedaldi, and Zisserman]{parkhi2015a}
Parkhi, O., Vedaldi, A., and Zisserman, A.
\newblock Deep face recognition.
\newblock pp.\  1--12. British Machine Vision Association, 2015.

\bibitem[Phillips et~al.(2005)Phillips, Flynn, Scruggs, Bowyer, Chang, Hoffman,
  Marques, Min, and Worek]{phillips2005overview}
Phillips, P.~J., Flynn, P.~J., Scruggs, T., Bowyer, K.~W., Chang, J., Hoffman,
  K., Marques, J., Min, J., and Worek, W.
\newblock Overview of the face recognition grand challenge.
\newblock In \emph{2005 IEEE computer society conference on computer vision and
  pattern recognition (CVPR'05)}, volume~1, pp.\  947--954. IEEE, 2005.

\bibitem[Ringrose(2019)]{ringrose2019law}
Ringrose, K.
\newblock Law enforcement's pairing of facial recognition technology with
  body-worn cameras escalates privacy concerns.
\newblock \emph{Va. L. Rev. Online}, 105:\penalty0 57, 2019.

\bibitem[Scherhag et~al.(2019)Scherhag, Rathgeb, Merkle, Breithaupt, and
  Busch]{scherhag2019face}
Scherhag, U., Rathgeb, C., Merkle, J., Breithaupt, R., and Busch, C.
\newblock Face recognition systems under morphing attacks: A survey.
\newblock \emph{IEEE Access}, 7:\penalty0 23012--23026, 2019.

\bibitem[Scheuerman et~al.(2019)Scheuerman, Paul, and
  Brubaker]{scheuerman2019computers}
Scheuerman, M.~K., Paul, J.~M., and Brubaker, J.~R.
\newblock How computers see gender: An evaluation of gender classification in
  commercial facial analysis services.
\newblock \emph{Proceedings of the ACM on Human-Computer Interaction},
  3\penalty0 (CSCW):\penalty0 1--33, 2019.

\bibitem[Schroff et~al.(2015)Schroff, Kalenichenko, and
  Philbin]{schroff2015facenet}
Schroff, F., Kalenichenko, D., and Philbin, J.
\newblock Facenet: A unified embedding for face recognition and clustering.
\newblock In \emph{Proceedings of the IEEE conference on computer vision and
  pattern recognition}, pp.\  815--823, 2015.

\bibitem[Schwartz(2012)]{schwartz2012chicago}
Schwartz, A.
\newblock Chicago's video surveillance cameras: A pervasive and poorly
  regulated threat to our privacy.
\newblock \emph{Nw. J. Tech. \& Intell. Prop.}, 11:\penalty0 ix, 2012.

\bibitem[Sharif et~al.(2016)Sharif, Bhagavatula, Bauer, and
  Reiter]{sharif2016accessorize}
Sharif, M., Bhagavatula, S., Bauer, L., and Reiter, M.~K.
\newblock Accessorize to a crime: Real and stealthy attacks on state-of-the-art
  face recognition.
\newblock In \emph{Proceedings of the 2016 acm sigsac conference on computer
  and communications security}, pp.\  1528--1540, 2016.

\bibitem[Smith(2019)]{smith2019more}
Smith, A.
\newblock More than half of us adults trust law enforcement to use facial
  recognition responsibly.
\newblock \emph{Pew Research Center}, 2019.

\bibitem[van Heek et~al.(2016)van Heek, Arning, and
  Ziefle]{van2016surveillance}
van Heek, J., Arning, K., and Ziefle, M.
\newblock The surveillance society: Which factors form public acceptance of
  surveillance technologies?
\newblock In \emph{Smart Cities, Green Technologies, and Intelligent Transport
  Systems}, pp.\  170--191. Springer, 2016.

\bibitem[Viola \& Jones(2004)Viola and Jones]{viola2004robust}
Viola, P. and Jones, M.~J.
\newblock Robust real-time face detection.
\newblock \emph{International journal of computer vision}, 57\penalty0
  (2):\penalty0 137--154, 2004.

\bibitem[Wang \& Kosinski(2018)Wang and Kosinski]{wang2018deep}
Wang, Y. and Kosinski, M.
\newblock Deep neural networks are more accurate than humans at detecting
  sexual orientation from facial images.
\newblock \emph{Journal of personality and social psychology}, 114\penalty0
  (2):\penalty0 246, 2018.

\bibitem[Wu \& Zhang(2016)Wu and Zhang]{wu2016automated}
Wu, X. and Zhang, X.
\newblock Automated inference on criminality using face images.
\newblock \emph{arXiv preprint arXiv:1611.04135}, pp.\  4038--4052, 2016.

\bibitem[Zhu et~al.(2015)Zhu, Lei, Yan, Yi, and Li]{zhu2015high}
Zhu, X., Lei, Z., Yan, J., Yi, D., and Li, S.~Z.
\newblock High-fidelity pose and expression normalization for face recognition
  in the wild.
\newblock In \emph{Proceedings of the IEEE Conference on Computer Vision and
  Pattern Recognition}, pp.\  787--796, 2015.

\end{thebibliography}
\bibliographystyle{icml2020}

\end{document}